# Evolutionary dynamics of cryptocurrency transaction networks: An empirical study

**Jiaqi Liang[1,2]\*, Linjing Li[1]\*, Daniel Zeng[1,2,3]**

**1** The State Key Laboratory of Management and Control for Complex Systems, Institute of Automation, Chinese Academy of Sciences, Beijing, China, **2** School of Computer and Control Engineering, University of Chinese Academy of Sciences, Beijing, China, **3** Department of Management of Information Systems, The University of Arizona, Tucson, AZ, United States of America

\* liangjiaqi2014@ia.ac.cn (JL); linjing.li@ia.ac.cn (LL)

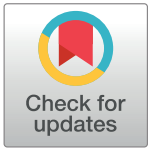

**Data Availability Statement:** All relevant data are available from the Harvard Dataverse database (doi: 10.7910/DVN/IO50XZ, 10.7910/DVN/XIXSPR, 10.7910/DVN/M9K50J).

**Funding:** This work was supported in part by the National Key Research and Development Program of China under Grant (Nos. 2017YFC0820105 to L. L., 2016QY02D0305 to D.Z., http://www.most.gov. cn/kjjh/), the National Natural Science Foundation of China (Nos. 71621002 to D.Z., U1435221 to L.L., http://www.nsfc.gov.cn/), and the Key Research Program of the Chinese Academy of Sciences under Grant (No. ZDRW-XH-2017-3 to

## Abstract

Cryptocurrency is a well-developed blockchain technology application that is currently a heated topic throughout the world. The public availability of transaction histories offers an opportunity to analyze and compare different cryptocurrencies. In this paper, we present a dynamic network analysis of three representative blockchain-based cryptocurrencies: Bitcoin, Ethereum, and Namecoin. By analyzing the accumulated network growth, we find that, unlike most other networks, these cryptocurrency networks do not always densify over time, and they are changing all the time with relatively low node and edge repetition ratios. Therefore, we then construct separate networks on a monthly basis, trace the changes of typical network characteristics (including degree distribution, degree assortativity, clustering coefficient, and the largest connected component) over time, and compare the three. We find that the degree distribution of these monthly transaction networks cannot be well fitted by the famous power-law distribution, at the same time, different currency still has different network properties, e.g., both Bitcoin and Ethereum networks are heavy-tailed with disassortative mixing, however, only the former can be treated as a small world. These network properties reflect the evolutionary characteristics and competitive power of these three cryptocurrencies and provide a foundation for future research.

## Introduction

Network analysis, such as those reported in [1–4], has attracted increasing attention in economics and finance since it provides further insights than traditional methods. Although a large volume of financial data, e.g., stock price, is available for network related research and analysis, information about transaction details is usually considered sensitive and not available for research. Cryptocurrency, where a continuously growing list of records stored in a chain is accessible, provides opportunities to analyze transaction networks in detail.

A cryptocurrency is a digital currency in which blockchain techniques are used to secure the transactions and control the generation of new units of currency (the so-called coins), operating independently without a central authority. Specifically, cryptocurrency relies on a private key to prove ownership and a public history of transactions to prevent double-spending [5]. Since Bitcoin [6], the first cryptocurrency, emerged in 2009, many other alternatives have emerged with modified rules of transaction and usage, e.g., Namecoin provides decentralized name registration [7], while Ethereum allows the automatic transfer of digital assets according to the so-called "smart contract" [8]. By introducing new types of assets and new transaction management methods, cryptocurrency has the potential to replace traditional fiat-currency. At the time of writing, there are over 900 currencies in the cryptocurrency market and the total market cap has exceeded $578 billion [9, 10]. Thus, it is the right time to investigate and compare them, so as to fully understand cryptocurrency and provide a foundation for future research.

Public availability of cryptocurrency transactions provides a basis for analyzing its transaction networks. For networks, especially the so-called complex networks, reported investigations mainly focus on descriptive statistics, network evolution, statistical mechanics of network topology and dynamics [11]. There are also studies on the robustness against failures and attacks, spreading processes and synchronization [12]. The descriptive statistics are majorly adopted to depict the behavior of Bitcoin users [13, 14]. In the field of Namecoin, Kalodner et al. [15] analyzed the uses and the transfers of the namespace, they also devised some principles on mechanism design. Relating to the evolution of networks, most networks encountered in practice have the tendency to densify over time [16], however, Bitcoin network densifies only in its first five years [17] and Namecoin network only densifies in the first year [18]. Motivated by empirical data, complex networks have some typical structure features, including small worlds, clustering, and degree distribution fitted by the power law. Baumann et al. [14] found that the Bitcoin system was a "small world" network and followed a scale-free distribution. Kondor et al. [19] further illustrated that the transaction networks are characterized by disassortative degree correlation in the trading phase, they applied linear preferential attachment to interpret it. Regarding research on multiple currencies, Anderson et al. [20] studied the characteristics of three representative cryptocurrencies separately, but no comparison of network characteristics was provided. Walsh et al. [21] identified eight key characteristics of system design and divided currencies into four archetypes, but there was no in-depth network analysis.

In this paper, we apply statistics and network analysis methods to explore the dynamic characteristics of three transaction networks. We download transaction data from the respective blockchain explorers. To the best of our knowledge, these are the largest datasets adopted in cryptocurrency analysis to date. We analyze the growth pattern of the accumulated network and find that unlike most networks, these cryptocurrency networks do not always densify over time. Then based on the datasets, we find that the monthly repetition ratios measured by either node or edge are relatively low. As such, studying the whole accumulated network, as done in most previous work [18, 19], is not the appropriate way to understand the network dynamics. Hence we focus on coining the dynamics through computing the values of typical network measures on a monthly basis, and make a comparison among the three networks.

The main contributions of our research are: 1) We find that the growth pattern of cryptocurrency transaction networks is different from that of most other networks reported in the literature in the way that they do not always follow neither the densification law nor the constant average degree assumption over time; 2) Monthly network, instead of accumulated network, is proposed as an appropriate object to understand the dynamics of the network; 3) we conduct the first empirical comparison among three representative cryptocurrency networks and point out the similarities and differences to help understand the peer-to-peer technology on a network level. Different from previous researches on complex networks, we find that the degree distribution of the cryptocurrency transaction networks cannot be well fitted by the famous pow-law distribution.

The remainder of this paper is organized as follows. In the next section, we provide our datasets, the necessary background to understand the transaction networks and our methodology used to analyze the networks. The Results section presents our findings for Bitcoin, Ethereum, and Namecoin networks. We offer our conclusions in the last section.

## Methodology

In this section, we first introduce the datasets used for analysis then explain how to construct a transaction network from corresponding dataset, the transaction network is the basis for the subsequent dynamic analysis. Finally, we introduce the measures used for the network analysis.

### Datasets

Among the complete list of cryptocurrencies, we choose three representatives for our analysis: Bitcoin, Namecoin, and Ethereum. Bitcoin is chosen as it is the first and by far the largest cryptocurrency; Namecoin is the first cryptocurrency that works as a decentralized domain name system; and Ethereum is the first cryptocurrency that supports "smart contract" and is also one of the most active cryptocurrencies. The data on transactions are from the blockchain explorers [22–24]. We believe, but cannot fully verify, that the data should be the same as what one could get as a cryptocurrency client. Even if there are tiny differences, they are likely to have only a negligible effect on our statistical results. We downloaded the complete list of transactions of each currency from its inception through 31 October 23:59:59 2017 UTC. A summary of the datasets is provided in Table 1.

### Transaction network

Blockchain is a distributed public ledger that records transactions ever verified in the network. It is implemented as a chain of blocks, each block containing a hash of the previous block up to the genesis block of the chain. And each block holds batches of valid transactions in the form of owner X transferring Y coins to payee Z. In the cryptocurrency system, payers and payees can create an unlimited number of addresses. A transaction in cryptocurrency system is a kind of regular bank transaction in the sense that it allows multiple sending addresses and multiple receiving addresses existing in a transaction.

Take the Bitcoin system as an example, it specifies how many Bitcoins are sent or received from an address, but there are no details of who sends how many Bitcoins to whom. Fig 1A shows an example of the transaction with two sending addresses and two receiving addresses

**Table 1. Summary of cryptocurrencies studied in this paper.**

| Cryptocurrency | Bitcoin | Namecoin | Ethereum |
|---|---|---|---|
| Time of genesis block | 2009-01-03 18:15:05 | 2011-04-17 00:26:41 | 2015-07-30 15:26:28 |
| #blocks | 492,558 | 368,347 | 4,467,004 |
| #transactions | 266,465,682 | 4,620,786 | 80,439,683 |
| #addresses | 363,937,999 | 6,906,323 | 10,241,475 |
| Size on disk | 71.72 GB | 1.34 GB | 3.95 GB |

Here "Time of genesis block" represents the time when the corresponding distributed ledger starts, "#" indicates the number of corresponding variables, and "Size on disk" represents the space occupied by the corresponding dataset on the disk.

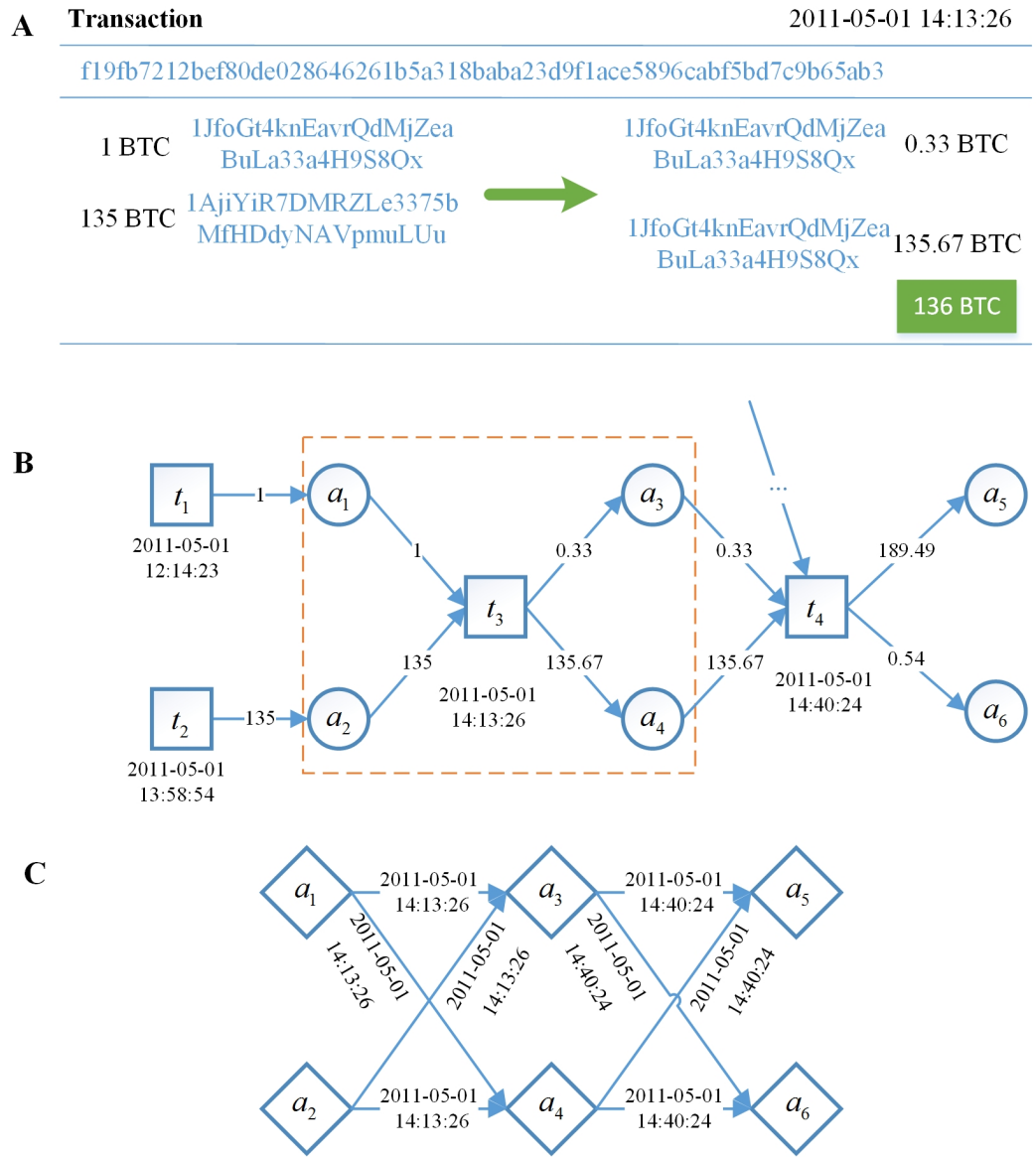

**Fig 1. Illustration of transaction network construction.** (A) An example of Bitcoin transaction details. (B) Example information extracted from Bitcoin transactions, and the information in the orange box correspond to the transaction in (A). (C) The Bitcoin transaction network as a directed graph.

which was added on the blockchain on May 1, 2011, and the relevant details can be queried on the corresponding crawling website through the identifier. Specifically, the time in the upper right corner indicates when the transaction was added to the blockchain, and the value on the first row is the transaction identifier, i.e., a hash value. Next, "1BTC" and its neighboring value (hash value of transaction address) denote that the address sent 1 Bitcoin. Therefore, Fig 1A shows a transaction that two sending addresses contribute 1 Bitcoin and 135 Bitcoins respectively, the two receiving addresses receive 135.67 Bitcoins (for payment) and 0.33 Bitcoins (maybe a transaction fee or a new address for remaining bitcoins). Fig 1B is an example information extracted from Bitcoin transactions where the value of the arrow represents corresponding value in Bitcoins that are flowing. Here $a_i$'s represent addresses and $t_i$'s denote transactions. $t_3$ is a transaction with two inputs ($a_1$ and $a_2$) and two outputs ($a_3$ and $a_4$). The transaction was added to the blockchain on May 1, 2011. $t_4$ is a transaction with five inputs and two outputs ($a_5$ and $a_6$) happened on the same day. note that two inputs (i.e. $a_3$ and $a_4$) of $t_4$ are connected to the aforementioned outputs of $t_3$. An input to a transaction is either the output of a previous transaction or incentives (including newly generated bitcoins and transaction fees) for users. Regarding the number of transaction inputs, it can be a single input from a previous larger transaction, or multiple inputs combining smaller amounts. For security purposes, a transaction may have multiple outputs: one for the transfer of the rest, if any, back to the sender, and the other is used for the payment.

Public availability of cryptocurrency transactions and the input-output relationship between transactions provide a basis for transaction network research. The transaction network represents the flow of cryptocurrency between addresses over time. In a transaction network, each node represents an address. Without the specific value of cryptocurrency flow from inputs and outputs, there is an edge with a timestamp between any sending address and receiving address existing in a transaction. For instance, Fig 1C shows the network constructed from transactions in Fig 1B.

## Network measures

In the first part of our analysis, several descriptive statistics are calculated to analyze the accumulated network growth. The number of edges and nodes are adopted to represent the network size. Many networks encountered in practice densify over time with the average degree increasing, which means the number of edges grow superlinearly with respect to the number of nodes. This property is quantified by $e(t) \sim n(t)^a$, where $e(t)$ and $n(t)$ denote the number of edges and nodes of the graph at time $t$ respectively, and $a > 0$ is an exponent indicating the network's tendency to become denser [16].

The second part of our analysis regards the network topology. Cryptocurrency networks vary as time goes by: nodes are added by creating new addresses and removed when they are no longer involved in any transaction, while new edges are created for transactions between two previously unconnected addresses. We first check the monthly repetition ratio, defined by Eq (1), to help find a valid investigation object:

$$Ratio_{Rep}^t = \frac{|E^t \cap E^{t-1}|}{|E^t \cup E^{t-1}|},\tag{1}$$

where $E^t$ refers to the set of edges or nodes in a network at time $t$, $\cap$ is the intersection operator and $\cup$ is the union operator as in ordinary set theory, and $|\cdot|$ gives the number of elements when applying to a specific set.

For the monthly networks, we further analyze the dynamic characteristics to investigate the topologic properties. We select four most representative measures for analysis, including degree distribution, degree assortativity, average clustering coefficient, and properties of the LCC. The network measures adopted are briefly introduced in the following.

**Degree distribution** captures the individual connectivity of nodes [11]. The in(out)-degree of a node represents the number of transactions it involves as output(input), and the degree distribution is the probability distribution of these degrees over the whole network. Empirically, observed complex networks tend to show a heavy-tailed distribution following a power-law distribution $p(k) \sim k^{-\gamma}$, where $k$ is the value of the degree and the coefficient $\gamma$ has been found to be the characteristic of a complex network [25].

**Degree assortativity** measures the node preference—that nodes with similar degrees tend to be connected to each other [26]. Its strength, expressed as the degree assortativity coefficient, denoted by $r$, is defined as:

$$r = \frac{M^{-1}\sum_i j_i k_i - \left[M^{-1}\sum_i \frac{1}{2}(j_i + k_i)\right]^2}{M^{-1}\sum_i \frac{1}{2}(j_i^2 + k_i^2) - \left[M^{-1}\sum_i \frac{1}{2}(j_i + k_i)\right]^2},$$ (2)

where $j_i$ and $k_i$ are the degrees of the nodes at the ends of the $i$-th edge, and $M$ is the number of edges.

**Clustering coefficient** represents the tendency of the nodes in a graph to cluster together, and the overall level of clustering is measured as the average of the local clustering coefficient of all nodes:

$$C = \frac{1}{N}\sum_v \frac{|\Delta_v|}{d_v(d_v - 1)/2},$$ (3)

where $|\Delta_v|$ denotes the number of triangles containing node $v$. To calculate $|\Delta_v|$, we ignore the directionality of the graph, and $d_v$ is the degree of node $v$ in the undirected graph. Watts and Strogatz [25] applied the clustering coefficient to discover small-world phenomenon within several networks.

**The largest connected component (LCC)** is a maximal subgraph in which any two nodes can be connected by a path. LCC is an important factor in understanding the network structure [11]. In this paper, we adopt relative size and the diameter of the LCC. The relative size is calculated by dividing the number of nodes that connect to the LCC by the number of nodes in the whole network. The diameter is the longest shortest path among all the nodes that form the LCC.

## Results

The analysis of cryptocurrency networks is conducted from three perspectives. In the first part, we explore the accumulated network growth. Then we select the appropriate investigation object for analysis. In the last part, we focus on analyzing the dynamics of the monthly networks and making comparisons. The analysis program is implemented in Python with the aid of powerlaw [27], Networkit [28], and statsmodel [29] packages.

### Accumulated network growth

In this part, we investigate the network growth from cryptocurrencies' inception till 31 October, 2017. For each month $m$, we construct a network using all transactions published up to month $m$. We analyze two aspects: network size (number of nodes and edges) and average degree.

The number of edges and nodes can be adopted to represent the size of the network, and they indicate the adoption rate and competitiveness of currency. As shown in Fig 2, the growth process can be divided into two phases.

- **Initial phase**. The system had low activity. Users just tried the currency experimentally and compared it with other currencies to find relative advantages. When a currency became more popular, more users would adopt it. Therefore, the network exhibited growing tendency with excessive fluctuations.

- **Trading phase**. With a certain number of adopters, growth slowed and did not change significantly. A reason is that the currency is constantly being accepted and rejected as a result of competition with other cryptocurrencies in the market.

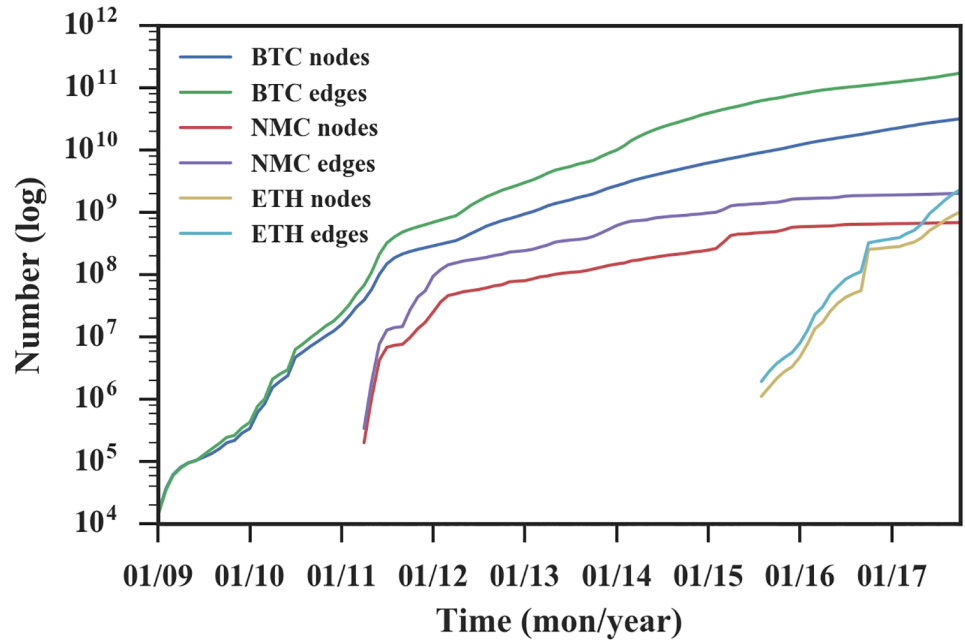

**Fig 2. The size of accumulated transaction networks with respect to various cryptocurrencies in log coordinate.** The number of nodes and edges are used to represent the size of networks. The three networks have similar growth pattern with rapid growth first and slower growth later.

When referring to the specific growth rates and duration time, Bitcoin grew over 10,000 times bigger in its first two-and-a-half years, while Namecoin and Ethereum grew over 100 times bigger in their first year. A reason for Bitcoin's long duration is that during 2009 to 2010, cryptocurrency was a new concept and Bitcoin was the only cryptocurrency in the market. All users who wanted to try cryptocurrency had to choose Bitcoin.

Then we investigate the average degree over time to find the network's tendency to become dense. Growth patterns in Fig 3 show the differences among the three networks. For Bitcoin, the average degree increased over time until September 2015. Subsequently, the decrease lasted for almost two years, probably because it had issues, such as hard to mine and large price fluctuations, and its competitor Ethereum offered a new option, "smart contract," for users interested in cryptocurrencies. Bitcoin has shown an increase since July 2017. For Namecoin, except for the increase in the initial phase, the average degree remained constant with some fluctuations due to competition among currencies. For Ethereum, the average degree continued increasing except for a decrease in October 2016. We suspect that the network instability is caused by a number of denial-of-service (DoS) attacks in late September and the two-stage "hard fork" to secure the network [30].

To gain more insight, we plot the number of nodes versus the number of edges for each cryptocurrency network on a logarithmic scale and fit a line reflecting the overall growth pattern of the network, as shown in Fig 4. The fitting parameters are shown in Table 2. For Bitcoin, the exponent is $a = 1.15$, which is clearly greater than 1, indicating a large deviation from linear growth with increasing average degree. For Ethereum and Namecoin, the exponent is close to 1, corresponding to the constant average degree over time. We also check the latest 1/3 of the data. Surprisingly, the Bitcoin network exponent is less than 1, the Ethereum network exponent is larger than 1, and the Namecoin exponent is close to 1, which coincides with the findings in Fig 3. The difference between the results of all data and the last 1/3 of the data indicates that the overall trend does not represent the real-time situation.

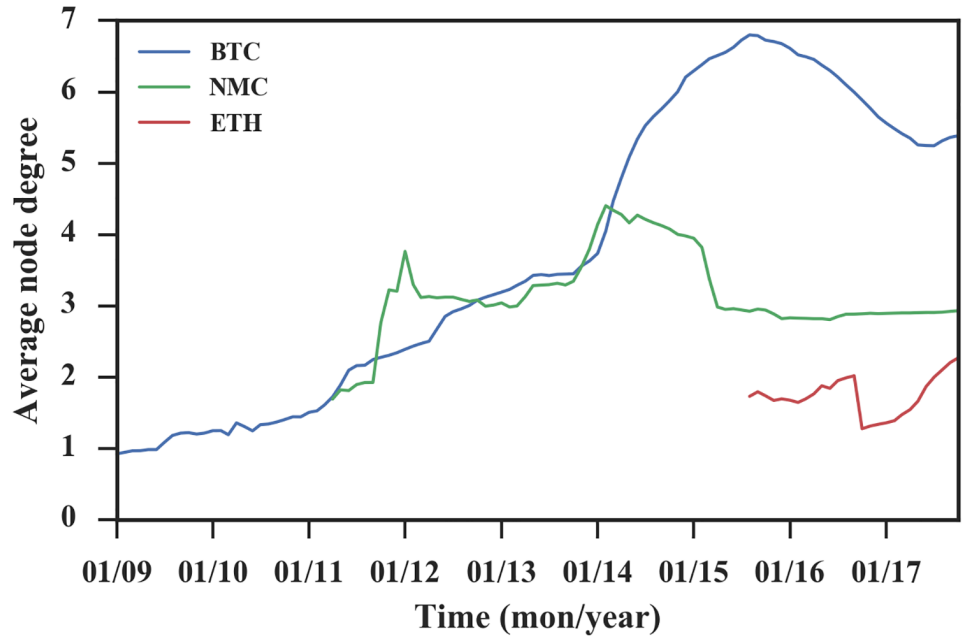

**Fig 3. The average node degree of accumulated networks over time.** The average degree of the three networks is not constant.

The above analysis on accumulated networks indicts that the cryptocurrency transaction networks do not always follow the densification law and the constant average degree assumption, which is different from most networks investigated in [16]. We must point out that there are several previous researches on cryptocurrency which have reported similar findings. Chang et al. [18] recognized that Namecoin only densifies in the first year while Holtz et al. [17] verified that Bitcoin densifies in the first five years. However, our conclusion is more valid and general since our conclusion is based on a quantitative analysis on three cryptocurrencies and our dataset covers a longer history.

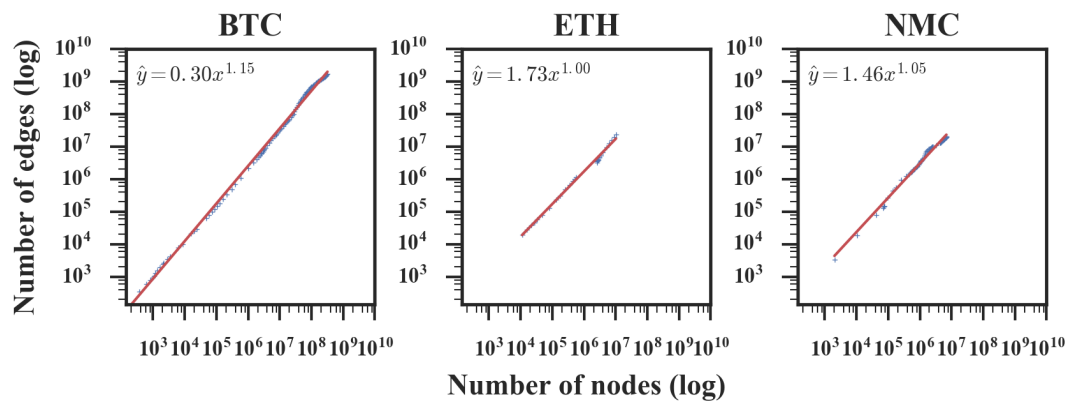

**Fig 4. The number of edges $e(t)$ versus the number of nodes $n(t)$ in accumulated transaction networks in log coordinate.** The red lines show fitted power-law distribution for the networks. In the figure's equation, $x$ represents the number of nodes and $\hat{y}$ represents the fitting number of edges, and the exponents are 1.15, 1.00, 1.05, respectively.

**Table 2. Fitting parameters of the power law.**

|  | total | | | lastest 1/3 of the data | | |
|---|---|---|---|---|---|---|
|  | *a* | $R^2$ | 95% confidence level | *a* | $R^2$ | 95% confidence level |
| BTC | 1.15 | 0.998 | [1.145, 1.163] | 0.86 | 0.987 | [0.830, 0.900] |
| ETH | 1.00 | 0.995 | [0.970, 1.029] | 1.38 | 0.998 | [1.332, 1.429] |
| NMC | 1.05 | 0.989 | [1.027, 1.078] | 0.99 | 0.982 | [0.938, 1.049] |

Here *a* is the exponent of $e(t) \sim n(t)^a$, and $R^2$ is the coefficient to measure the goodness of fit. It ranges from 0 to 1, the better the power law fits the data, the closer the value of $R^2$ is to 1.

Why do the cryptocurrency networks not obey the densification law? Security is the most probable explanation. In cryptocurrency system, to securely receive, store, and send coins, a user can spread his coins in multiple wallets, corresponding to multiple nodes in the network, to reduce risks. Therefore, in a transaction network, one user may have multiple nodes corresponding to multiple addresses. While in other real networks, a user usually has only one node.

## The object for dynamic analysis

Since the nodes and edges of the networks are changing all the time, we checked the monthly repetition ratio as shown in Fig 5. As to the nodes, Bitcoin and Namecoin have repetition ratio less than 0.1, while the value of Ethereum is less than 0.25. As to the edges, Bitcoin and Namecoin have repetition ratio less than 0.1, and the value of Ethereum is less than 0.2. Thus, after the initial phase, both the node and edge repetition ratio reach relatively low values, indicating that a lot of nodes and edges do not survive from one time window to the next and network reconfiguration takes place all the time. The low survival ratio of both nodes and edges can be illustrated by the aforementioned security reason as well. Specifically, in order to enhance security, users may constantly change their addresses. These surviving addresses may be the addresses of fixed payees, such as donors and miners, and these addresses are used to receive cryptocurrencies which are inconvenient to replace.

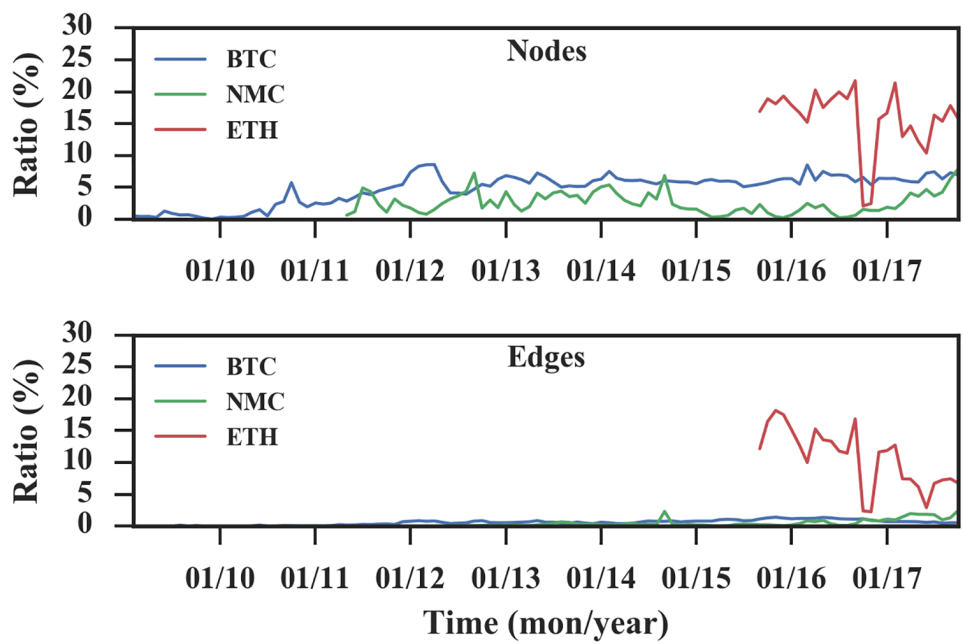

**Fig 5. Monthly repetition ratios over time.** (A) The ratios of edges. (B) The ratios of nodes. After the initial phase, all ratios reach relatively low values.

Due to the above characteristics of the cryptocurrency transaction networks, it is better to analyze the transaction networks in separate time intervals, rather than in the accumulated manner as done by most of the previous works. In the following of this paper, we set the time interval as one month and construct the monthly transaction networks to understand the dynamics of the transaction networks.

## Monthly network analysis

In this part, we present our main results on dynamic characteristics of cryptocurrency based on the monthly networks.

**Degree distribution.** We first check whether the degree distribution of the three representatives can be fitted by the power law. In the case of cryptocurrency networks, often the initial, small values of the data do not follow the power-law distribution, thus we ignore these data when fitting. Further, we use the Kolmogorov-Smirnov (KS) test to assess the goodness-of-fit. We find that almost all degree distributions cannot be accepted as the power law strictly under the 95% confidence level. However, the degree distribution is still a clear heavy-tailed distribution, which means that the majority of addresses have low degrees, while small but not negligible addresses have relatively high degrees. As shown in Fig 6, we divide the phases as follows.

- The number of adoption users is small, and there exist large errors in the fitting. Specifically, Bitcoin has a longer duration for the reason discussed in the "Accumulated network analysis" section.

- With a certain number of adoption users, the data are approximately fitted by the power law, though the acceptance rate using the KS test of power-law fit on the degree distribution is low. And the exponent fluctuates within a certain range. Specifically, the ranges of the exponent $\gamma$ are: [2.0, 3.0] for Bitcoin, [1.5, 3.0] for Ethereum, and [1.5, 3.5] for Namecoin. Note that the coefficient for the power law typically lies in the range [2, 3] as reported in [31].

- Due to fierce competition with other currencies, the range of data that satisfies the power-law distribution narrows. During and after the transition phase, different currencies have different features as follows: for Bitcoin and Ethereum, after the transition stage, the data are again approximately fitted by the power law and the exponents do not change significantly; for Namecoin, in the transition stage, there exists a phenomenon that the number of nodes with large degree is large too, thus it does not fit the power law.

Our analysis suggests that when adoption users reach a certain amount, the distribution approximately fits with the power law. However, under market competition, the scale of Namecoin network' nodes stabilize at a relatively small level of ten thousand, while the other two networks have millions of nodes. At the same time, due to the specific function of domain registration, there are some enthusiasts who insist on using Namecoin, leading to the phenomenon that the number of nodes with large degree is large too. In summary, under the effect of market competition, failed currencies do not fit well with the power law, while successful currencies approximately fit with the power law with fixed exponents.

**Degree assortativity.** We use the in-assortativity $r(in, in)$ and out-assortativity $r(out, out)$ to further investigate how the nodes are mixing by the degree in the network. A positive value for $r$ (assortative mixing) indicates that high-degree nodes are preferentially attached to other high-degree nodes. A negative value for $r$ indicates disassortative mixing, i.e., high-degree nodes are prone to connect to low-degree ones. Finally, $r = 0$ (neutral mixing) indicates the network is non-assortative.

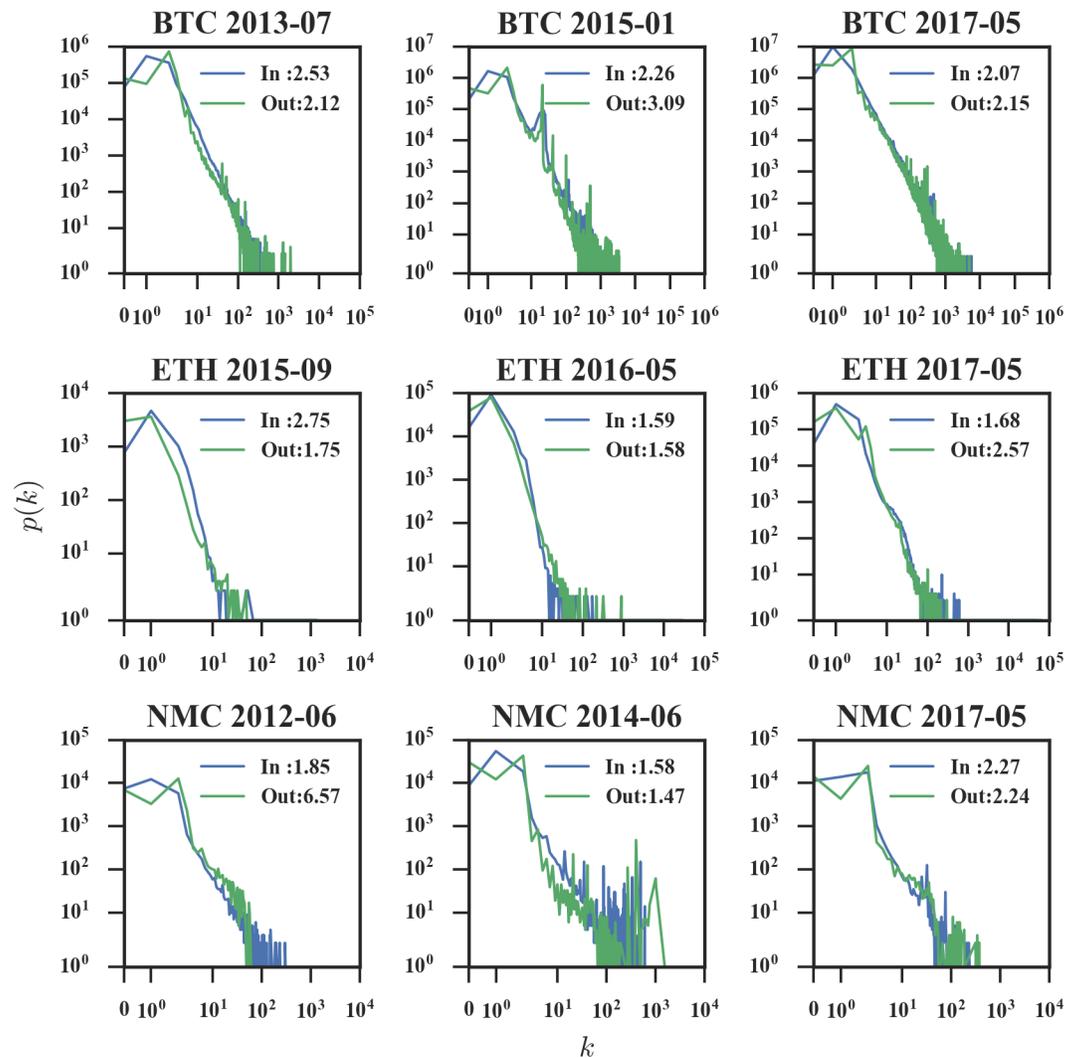

**Fig 6. Samples of degree distributions of monthly networks.** Data are sampled from Bitcoin (top row), Ethereum (middle row), and Namecoin (bottom row). Example data for power-law fitting are approximate fit (first column), poor fit (medium column), and inconsistent fit (last column). The legends show the fitting exponent $\gamma$ in $p(k) \sim k^{-\gamma}$ with respect to indegree and outdegree distribution.

As shown in Fig 7, except for the initial phase, the ranges of the in-assortativity and out-assortativity are [-0.05, 0] for Bitcoin and [-0.1, 0] for Ethereum, which suggests that the two networks are disassortative. The coefficients of Namecoin stay in the range of [-0.1, 0.1], making it difficult to judge its degree assortativity. In general, small values of $r$ are hard to interpret, thus we measure the quantity $\langle k_{nn} \rangle = \sum_{k'} k' P_c(k'|k)$, i.e., the average degree of nearest neighbors of nodes with degree $k$, for the in-degree and out-degree of the last month (October 2017).

In networks without degree correlations, the degrees of connected nodes do not depend on each other. Therefore for such networks, we expect the $\langle k_{nn} \rangle$ of the in-degree and out-degree should be constant. Regarding the last month's cryptocurrency transaction network, as can be seen from Fig 8, we find that for the Bitcoin and Ethereum networks, $\langle k_{nn} \rangle$ is a decreasing function, which indicates that nodes with high degree are prone to connect with low-degree nodes, thus the networks are disassortative. For the Namecoin network, the curve is nearly constant, which means that the degrees of connected nodes do not rely on each other, so the Namecoin network is non-assortative. A possible reason is that for highly heterogeneous (scale-free) networks, the maximum entropy principle leads to disassortativity [32]. Thus the cause of the difference in assortativity is also the market competition.

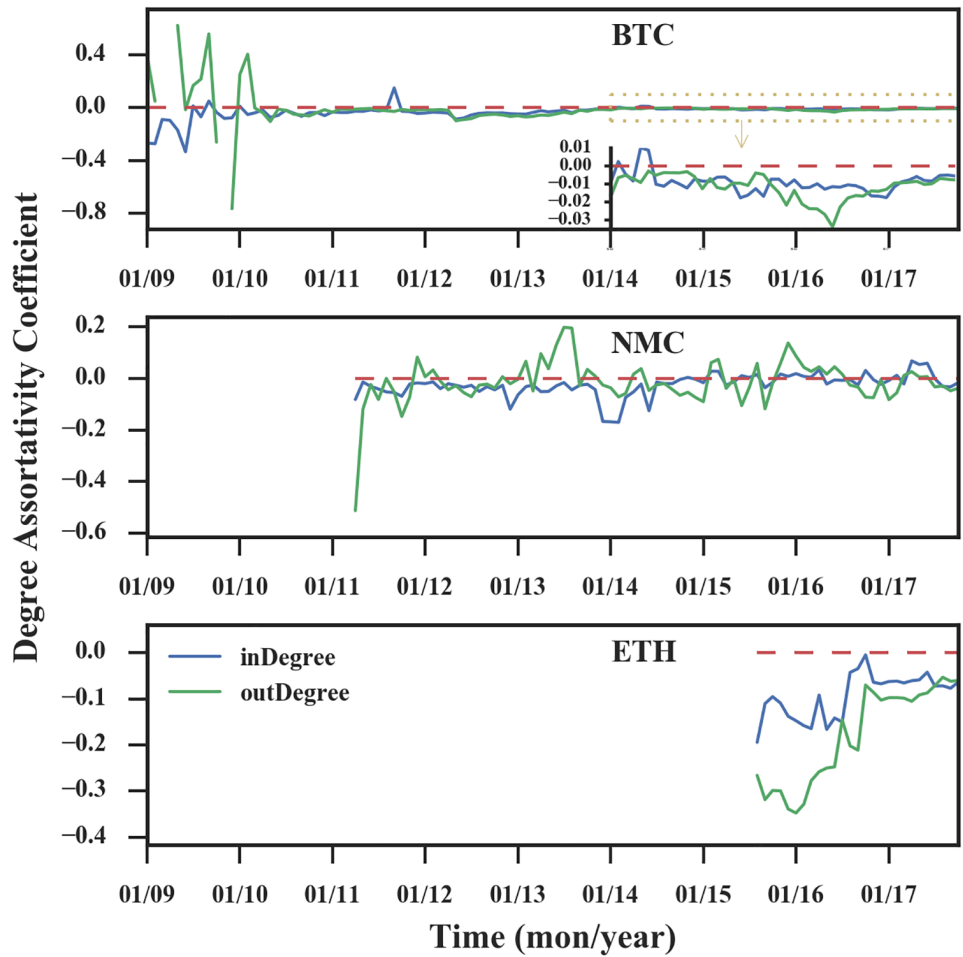

**Fig 7. Evolution of the degree assortativity.** In the figure of Bitcoin, we magnify the y-axis of the data in the yellow box and display it in the bottom right corner. After the initial phase, the coefficients of Bitcoin and Ethereum are negative, and the coefficient of Namecoin converges to a certain range near 0.

In summary, based on the analysis of the networks' in/out-assortativity and $\langle k_{nn} \rangle$ of in/out-degree, we find that Bitcoin and Ethereum networks are disassortative, while the Namecoin network is non-assortative, which is consistent with the observation that the degree distributions are heavy-tailed for the Bitcoin and Ethereum networks.

**Average clustering coefficient.** In order to find the evidence for a small-world network, we further compare the average clustering coefficients of networks to a random network with the same degree sequence [33]. In cryptocurrency networks, small-world means the currencies can be transferred between most nodes in the network by a small number of hops or steps if they want.

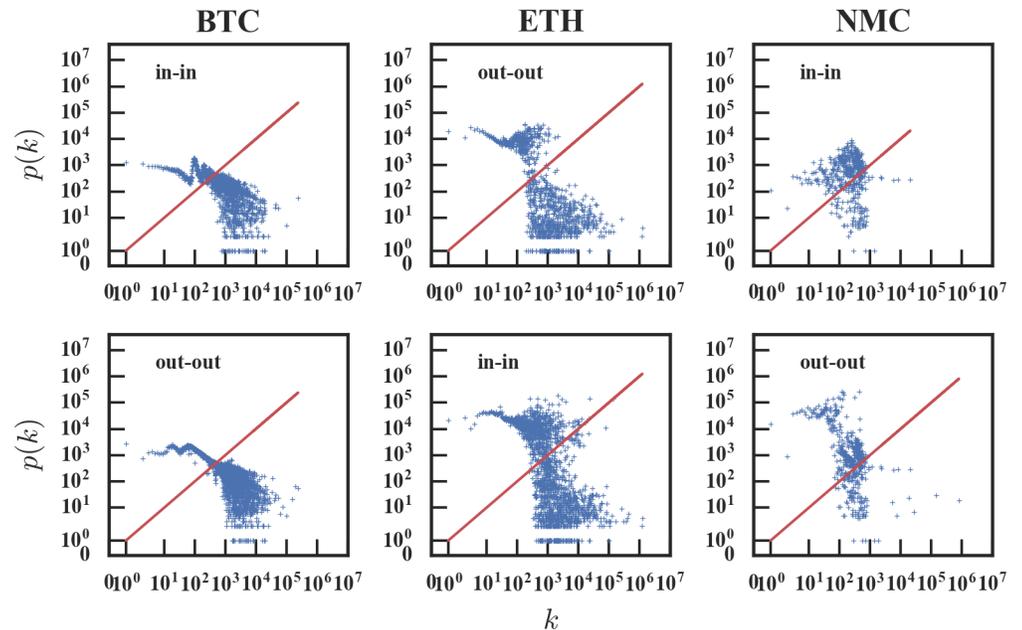

**Fig 8. A comparison of the average degree of nodes' neighbors and the degree of nodes.** The red line indicates the degree of the node and the average degree of the node's neighbors is equal. In networks without degree correlations, the $\langle k_{nn} \rangle$ is constant. However, for Bitcoin and Ethereum, $\langle k_{nn} \rangle$ is a decreasing function, while for Namecoin $\langle k_{nn} \rangle$ is not.

As shown in Fig 9, for these three networks, in the initial phase, there is no significant difference between the clustering coefficient of the cryptocurrency network and the coefficient of the random network with the same degree sequence, and even sometimes the value of the random network is larger than the value of transaction network. In the latter phase, the three networks behave differently. Specifically, for Bitcoin, the clustering coefficient reaches a stationary value around $C \approx 0.05$, which is still higher than the clustering coefficient of random networks with the same degree sequence ($C \approx 0.0045$). As a developing currency, Ethereum network was abnormal in middle stage (from August to December 2016). While for Namecoin, the coefficients are not always higher than the coefficients of random networks.

The phenomenon at the initial stage maybe results from transactions taking place between addresses belonging to a few enthusiasts who try to play the system by moving cryptocurrencies between their addresses. The possible reason for the later phase of Bitcoin is that it is disassortative, which means newly added nodes tend to attach to high degree nodes, resulting the nodes tend to cluster together and form a small world. Ethereum's abnormalities in 2016 were caused by the network instability. And Namecoin network is non-assortative, that is, there is no correlation between pairs of linked nodes, thus the network does not exhibit this property.

**Properties of the largest connected component (LCC).** Last but not least, we measure the relative size and diameter of the LCC in the transaction network (Fig 10).

For the Bitcoin network, after the initial phase, the LCC connects about 60% of the nodes in the network. For the Ethereum network, the percentage of LCC connecting nodes rises with a fluctuation and most recently connects about 40% of the nodes. And for the Namecoin network, the LCC connects less than 5% of the nodes in the network. Therefore, the relative sizes of the LCC of Bitcoin and Ethereum are relatively large, while the size of Namecoin network is relatively small. This result is consistent with our previous finding that the Bitcoin and Ethereum networks are disassortative mixing in the ways that the new nodes with lower degrees tend to connect to the nodes with higher degrees and vice versa, while the Namecoin network is non-assortative as there is no correlation between nodes.

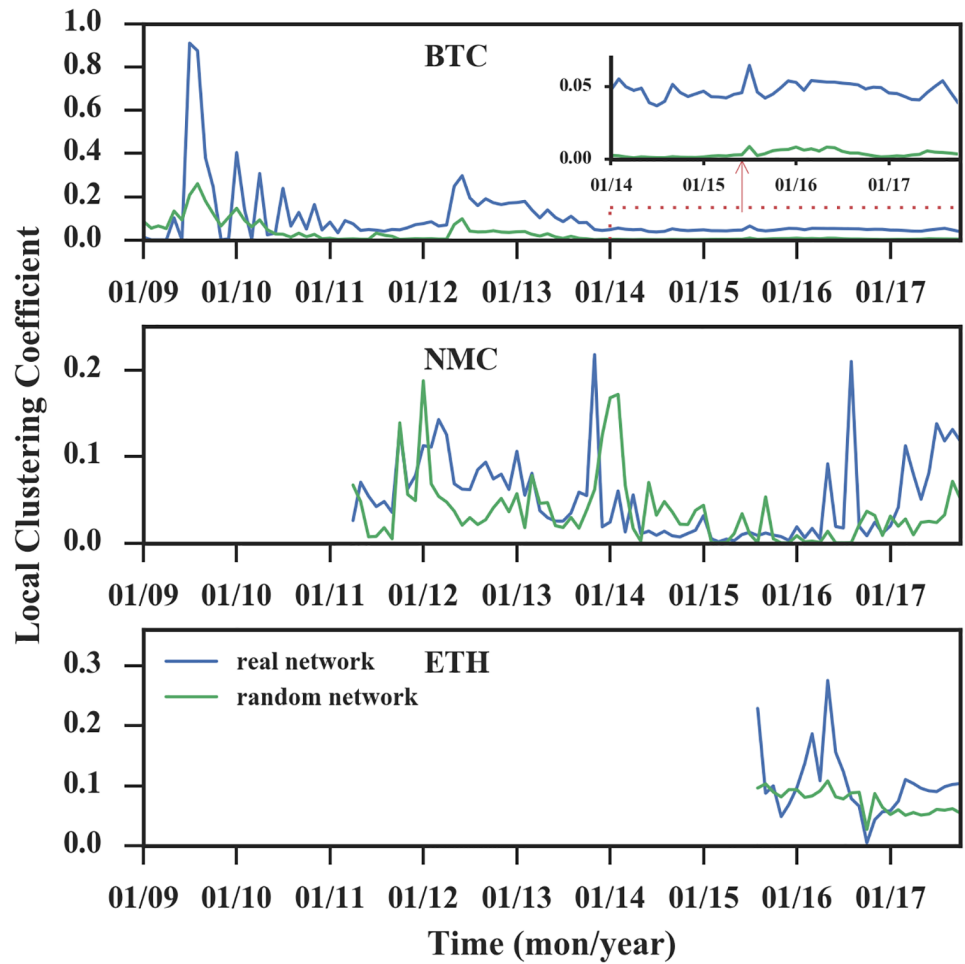

**Fig 9. Evolution of clustering coefficients.** If the average clustering coefficient of a network is rather higher than a random network with the same degree sequence, the network is a small-world network. In the figure of Bitcoin, we magnify the y-axis of the data in the red box and display it in the upper right corner. We find only Bitcoin exhibits this feature.

The diameter of the Bitcoin LCC is around 100, indicating inefficient system transfer, which is possibly the result of anonymous users trying to hide their identity by moving their own bitcoins as reported in [14]. The Ethereum LCC diameter is gradually increasing, possibly because the network is in its developing phase. The diameter of the Namecoin LCC fluctuates, which may be caused by the competition with other cryptocurrencies. Thus, during the period of our analysis, the LCCs of these three networks do not have a sign of shrinking diameter, and the possible reason may be the same as the reason that Bitcoin's LCC has a larger diameter.

## Discussion and conclusion

This paper analyzed the dynamic characteristics of the transaction networks of three representative cryptocurrencies: Bitcoin, Ethereum, and Namecoin. We first analyzed the growth of the transaction networks as they reflect the relative competitiveness of the cryptocurrencies under investigation. By analyzing the accumulated network growth, we find similar growth patterns: they all grow with large fluctuations in the initial phase then their growth slows down and fluctuates within a narrow range. However, unlike most networks reported in the literature, the cryptocurrency transaction networks do not always densify over time. This phenomenon is possibly due to the anonymity of the cryptocurrencies, where a user can create multiple addresses to receive, store, and send cryptocurrencies.

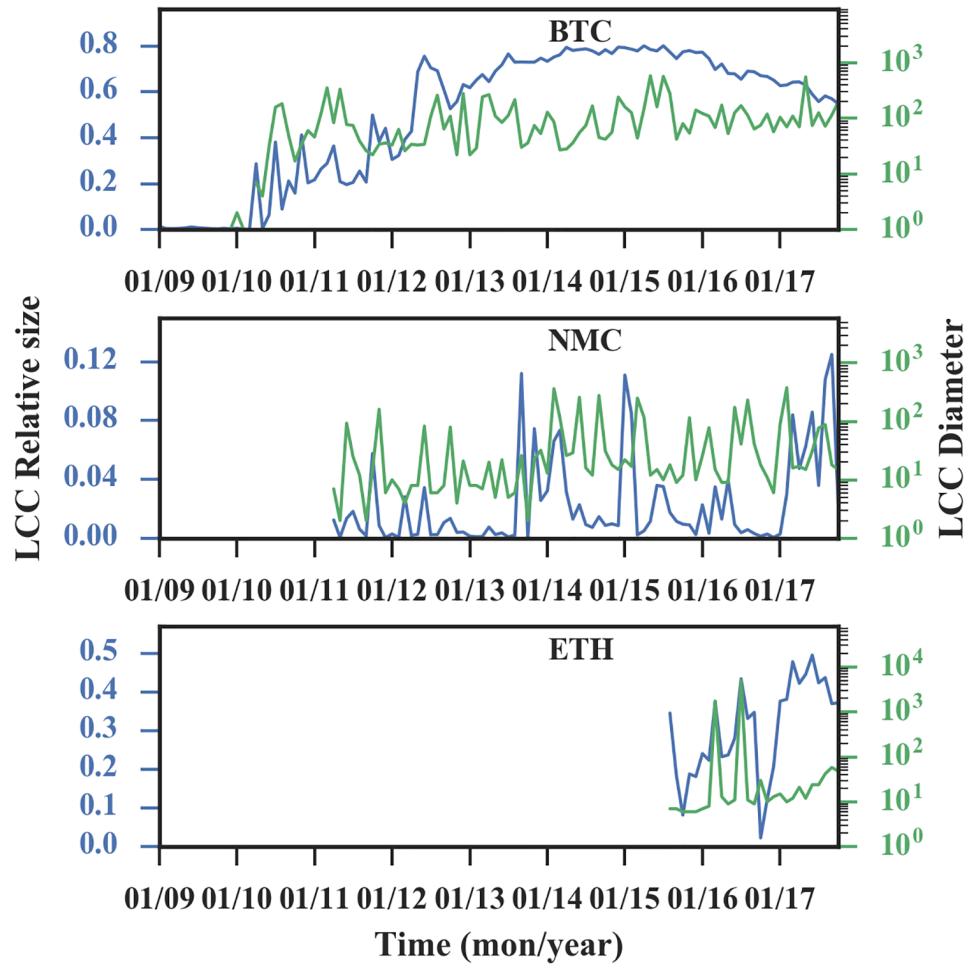

**Fig 10. The properties of the LCC.** The relative size (blue line) is the proportion of LCC nodes in all nodes, and the diameter (green line) reflects the connectivity of the LCC. Later stage, the relative sizes of the three networks are 60%, 40% and 5% respectively, and the diameter of BTC is 100, while the other two are in fluctuations.

Through computing the repetition ratio, we found that the overall accumulated network is not an applicable research object to investigate cryptocurrency properties. We thus conducted a monthly analysis on typical network measures and obtained the following insights on these three currencies. 1) Both Bitcoin and Ethereum networks may converge to heavy-tailed distribution in the long run, however, their degree distribution can only be approximately fitted by the power-law distribution. For Namecoin, its degree distribution cannot be fitted by the power-law distribution. 2) Bitcoin and Ethereum networks exhibit disassortative mixing, that is, newly added nodes tend to connect to nodes with higher degree. 3) Only the Bitcoin network is a small-world network according to the analysis of clustering coefficients. 4) Bitcoin and Ethereum's LCCs contain a relatively large proportion of nodes, however, the anonymity of Bitcoin results in a relatively large diameter.

The causes of these differences might be the original design ideas and user adoption of the cryptocurrencies. Since Bitcoin is the oldest and the most dominant cryptocurrency in the market, it was the unique choice for enthusiastic users, especially in the early days. At the same time, price volatility is another reason to attract users to Bitcoin by treating it as an investment alternative. Thus it is reasonable that Bitcoin network is heavy-tailed and organized as a small world. Ethereum, as a younger cryptocurrency, allows developers to write their own programs by replacing Bitcoin's more restrictive language with "smart contract", which attracts a great deal of user attention after its emergence. As a developing cryptocurrency, its network is heavy-tailed, but not a small world. The original design idea of Namecoin is to create a decentralized domain system, in which users can pay Namecoin to register and update the names of their domain. However, there are some other competitors in the market, say EmerCoin and NXT, which provide similar functionality. And this may be treated as the cause of its volatility characteristics.

Our findings suggest that these network properties reflect the evolutionary characteristics and competitive power of different cryptocurrencies. In the future, we will relate these transaction network properties with the currency characteristics to guide the design of digital financial products, policy regulations, and legislation.

## Author Contributions

**Conceptualization:** Jiaqi Liang, Linjing Li.

**Data curation:** Jiaqi Liang.

**Formal analysis:** Jiaqi Liang, Linjing Li.

**Funding acquisition:** Daniel Zeng.

**Investigation:** Jiaqi Liang, Linjing Li.

**Methodology:** Jiaqi Liang.

**Project administration:** Linjing Li.

**Resources:** Daniel Zeng.

**Software:** Jiaqi Liang.

**Validation:** Jiaqi Liang, Linjing Li.

**Visualization:** Jiaqi Liang.

**Writing – original draft:** Jiaqi Liang.

**Writing – review & editing:** Linjing Li, Daniel Zeng.

## References

1. Cimini G, Squartini T, Garlaschelli D, Gabrielli A. Systemic risk analysis on reconstructed economic and financial networks. Scientific Reports. 2015; 5:15758. https://doi.org/10.1038/srep15758 PMID: 26507849

2. Diebold FX, Yılmaz K. On the network topology of variance decompositions: Measuring the connectedness of financial firms. Journal of Econometrics. 2014; 182(1):119–134. https://doi.org/10.1016/j.jeconom.2014.04.012

3. Kristoufek L. What are the main drivers of the Bitcoin price? Evidence from wavelet coherence analysis. PLOS ONE. 2015; 10(4):e0123923. https://doi.org/10.1371/journal.pone.0123923 PMID: 25874694

4.  Schweitzer F, Fagiolo G, Sornette D, Vega-Redondo F, Vespignani A, White DR. Economic networks: The new challenges. Science. 2009; 325(5939):422–425. https://doi.org/10.1126/science.1173644 PMID: 19628858

5.  Narayanan A, Bonneau J, Felten E, Miller A, Goldfeder S. Bitcoin and Cryptocurrency Technologies: A Comprehensive Introduction.  Princeton University Press; 2016.

6.  Nakamoto S. Bitcoin: A peer-to-peer electronic cash system. 2008 [cited 2018 Jan 19]. Available from: https://bitcoin.org/bitcoin.pdf.

7.  Loibl A, Naab J. Namecoin. In: Proceedings of the Seminars Future Internet (FI) and Innovative Internet Technologies and Mobile Communications (IITM). 2014. p. 107–113. Available from: https://www.net. in.tum.de/fileadmin/TUM/NET/NET-2014-08-1/NET-2014-08-1_14.pdf.

8.  Buterin V. A next-generation smart contract and decentralized application platform. 2014 [cited 2018 Jan 19]. Available from: https://github.com/ethereum/wiki/wiki/White-Paper.

9.  Coin Market Cap [Internet]. Cryptocurrency Market Capitalizations. Available from: http://coinmarketcap.com/currencies/views/all/ Cited 19 Jan 2018.

10.  McCrank J,Shumaker L and Mazzilli M. Factbox: Bitcoin futures contracts at CME and Cboe. Reuters. 2017 Dec 16 [cited 2018 Jan 19]. Available from: https://www.reuters.com/article/us-bitcoin-futures-contracts-factbox/factbox-bitcoin-futures-contracts-at-cme-and-cboe-idUSKBN1E92IR.

11.  Albert R, Barabási AL. Statistical mechanics of complex networks. Reviews of modern physics. 2002; 74(1):47–97. https://doi.org/10.1103/RevModPhys.74.47

12.  Boccaletti S, Latora V, Moreno Y, Chavez M, Hwang DU. Complex networks: Structure and dynamics. Physics Reports. 2006; 424(4):175–308. https://doi.org/10.1016/j.physrep.2005.10.009

13.  Ron D, Shamir A. Quantitative Analysis of the Full Bitcoin Transaction Graph. In: Sadeghi AR, editor. Financial Cryptography and Data Security; 2013 Apr 1-5;  Okinawa, Japan. Berlin:  Springer; 2013. p. 6–24.

14.  Baumann A, Fabian B, Lischke M. Exploring the Bitcoin Network. In: Proceedings of the 10th International Conference on Web Information Systems and Technologies; 2014 Apr 3-5; Barcelona, Spain. Springer; 2014. p. 369-374.

15.  Kalodner H, Carlsten M, Ellenbogen P, Bonneau J, Narayanan A. An empirical study of Namecoin and lessons for decentralized namespace design. In: Workshop on the Economics of Information Security; 2015 Jun 22-23; Delft, Netherlands. Citeseer. 2015.

16.  Leskovec J, Kleinberg J, Faloutsos C. Graphs over time: densification laws, shrinking diameters and possible explanations. In: Proceedings of the eleventh ACM SIGKDD international conference on Knowledge discovery in data mining; 2005 Aug 21-24; New York, USA. ACM; 2005. p. 177-187.

17.  Fortuna J, Holtz B, Neff J. Evolutionary Structural Analysis of the Bitcoin Network. 2013 [cited 2018 Jan 19]. Available from: https://pdfs.semanticscholar.org/35b2/cac7bb85b6051f80b9eacfe292435c84a5c0.pdf.

18.  Chang TH, Svetinovic D. Data Analysis of Digital Currency Networks: Namecoin Case Study. In: 2016 21st International Conference on Engineering of Complex Computer Systems (ICECCS) Nov 6-8; Dubai, United Arab Emirates. IEEE; 2016. p. 122-125.

19.  Kondor D, Pósfai M, Csabai I, Vattay G. Do the rich get richer? An empirical analysis of the Bitcoin transaction network. PLOS ONE. 2014; 9(2):e86197. https://doi.org/10.1371/journal.pone.0086197 PMID: 24505257

20.  Anderson L, Holz R, Ponomarev A, Rimba P, Weber I. New kids on the block: an analysis of modern blockchains; 2016. Preprint. Available from: arXiv:1606.06530. Cited 19 Jan 2018.

21.  Walsh C, OReilly P, Gleasure R, Feller J, Li S, Cristoforo J, et al. New kid on the block: a strategic archetypes approach to understanding the Blockchain. In: Proceedings of the International Conference on Information Systems; 2016 Dec 11-14; Dublin, Ireland. Association for Information Systems; 2016.

22.  Bitcoin Block Explorer—Blockchain [Internet]. Available from: https://blockchain.info/. Cited 19 Jan 2018.

23.  Ethereum BlockChain Explorer and Search [Internet]. Available: https://etherscan.io/. Cited 19 Jan 2018.

24.  Namecoin Explorer [Internet]. Available from: https://bitinfocharts.com/de/namecoin/explorer/. Cited 19 Jan 2018.

25.  Watts DJ, Strogatz SH. Collective dynamics of 'small-world' networks. Nature. 1998; 393(6684):440–442. https://doi.org/10.1038/30918 PMID: 9623998

26.  Newman ME. Assortative mixing in networks. Physical review letters. 2002; 89(20):208701. https://doi.org/10.1103/PhysRevLett.89.208701 PMID: 12443515

27. Alstott J, Bullmore E, Plenz D. powerlaw: a Python package for analysis of heavy-tailed distributions. PLOS ONE. 2014; 9(1):e85777. https://doi.org/10.1371/journal.pone.0085777 PMID: 24489671

28. Staudt CL, Sazonovs A, Meyerhenke H. NetworKit: A tool suite for large-scale complex network analysis. Network Science. 2016; 4(4):508–530. https://doi.org/10.1017/nws.2016.20

29. Seabold S, Perktold J. Statsmodels: Econometric and statistical modeling with python. In: Proceedings of the 9th Python in Science Conference; 2010 Jun 28-30; Austin, Texas. SciPy society Austin; 2010.

30. Castor A. Ethereum's Hard Fork Drags Classic Along with it. 2016 Oct 17 [cited 2018 Jan 19]. In: BTCMANAGER.COM [Internet]. Available from: https://btcmanager.com/imminent-hard-fork-aligns-with-ethereum-classic-principles/.

31. Clauset A, Shalizi CR, Newman ME. Power-law distributions in empirical data. SIAM review. 2009; 51(4):661–703. https://doi.org/10.1137/070710111

32. Johnson S, Torres JJ, Marro J, Munoz MA. Entropic origin of disassortativity in complex networks. Physical review letters. 2010; 104(10):108702. https://doi.org/10.1103/PhysRevLett.104.108702 PMID: 20366458

33. Aiello W, Chung F, Lu L. A random graph model for massive graphs. In: Proceedings of the Thirty-second Annual ACM Symposium on Theory of Computing; 2000 May 21-23; New York, USA. ACM; 2000. p. 171-180.